# Shared Causal Paths underlying Alzheimer's dementia and Type 2 Diabetes

Zixin Hu[1], Rong Jiao[2], Jiucun Wang[1,3], Panpan Wang[1], Yun Zhu[4], Jinying Zhao[4], Phil De Jager[5], David A Bennett[6], Li Jin[1,3] and Momiao Xiong[2*]

**Background:** Although Alzheimer's disease (AD) is a central nervous system disease and type 2 diabetes mellitus (T2DM) is a metabolic disorder, an increasing number of genetic epidemiological studies show clear link between AD and T2DM. The current approach to uncovering the shared pathways between AD and T2DM involves association analysis; however, such analyses lack power to discover the mechanisms of the diseases.

**Methods:** We develop novel statistical methods to shift the current paradigm of genetic analysis from association analysis to deep causal inference for uncovering the shared mechanisms between AD and T2DM, and develop pipelines to infer multilevel omics causal networks which lead to shifting the current paradigm of genetic analysis from genetic analysis alone to integrated causal genomic, epigenomic, transcriptional and phenotypic data analysis. To discover common causal paths from genetic variants to AD and T2DM, we also develop algorithms that can automatically search the causal paths from genetic variants to diseases and

**Results:** The proposed methods and algorithms are applied to ROSMAP dataset with 432 individuals who simultaneously had genotype, RNA-seq, DNA methylation and some phenotypes. We construct multi-omics causal networks and identify 13 shared causal genes, 16 shared causal pathways between AD and T2DM, and 754 gene expression and 101 gene methylation nodes that were connected to both AD and T2DM in multi-omics causal networks.

**Conclusions:** The results of application of the proposed pipelines for identifying causal paths to real data analysis of AD and T2DM provided strong evidence to support the link between AD and T2DM and unraveled causal mechanism to explain this link.

[1]State Key Laboratory of Genetic Engineering and Innovation Center of Genetics and Development, School of Life Sciences, Fudan University, Shanghai, China.

[2]Department of Biostatistics and Data Science, School of Public Health, University of Texas Health Science Center at Houston, Houston, Texas, USA.

[3]Human Phenome Institute, Fudan University, Shanghai, China.

[4]Department of Epidemiology, University of Florida, Florida, USA

[5] Center for Translational & Computational Neuroimmunology, Department of Neurology, Columbia University Medical Center, New York, 10033, USA

[6]Rush Alzheimer's Disease Center, Rush University Medical Center, Chicago, IL 60612, USA

## Background

Although Alzheimer's dementia is a central nervous system disease and type 2 diabetes mellitus (T2DM) is a metabolic disorder, an increasing number of epidemiological and genetic epidemiological studies show clear link between Alzheimer's dementia and T2DM. Alzheimer's dementia with great economic, political and social consequences is a progressive, irreversible degenerative disease of the brain and is the most common cause of dementia due to the gradual accumulation of amyloid-beta ($A\beta$) and twisting of tau protein [1, 2], and other common brain pathologies [3]. Alzheimer's dementia is also involved in inflammation and oxidative address and exhibits memory loss and cognitive dysfunction [4, 5].

Two mechanisms underlying T2DM are insulin resistance and insufficient insulin secretion from pancreatic $\beta$-cells [4]. T2DM patients are unable to process insulin signaling correctly that lead to the insulin-resistant. In response to insulin resistance, pancreatic $\beta$-cells increase insulin production. However, when pancreatic $\beta$-cells gradually lose function; insulin production cannot be increased to maintain normal glucose levels. The brain is a target organ for insulin [6]. Insulin signaling plays an important role in the organization and function of the brain and impaired insulin signaling induces an overactivation of GSK-3 kinase, increases tau phosphorylation, alters tau modification and causes neurofibrillary degeneration [7]. T2DM also suffer from mild to severe nervous system damage. Persistent blood glucose increases may impair blood flow to the brain [8].

Prior work in ROSMP found an association of T2DM with incident Alzheimer's dementia and rate of cognitive decline [9]. However, we did not find an association with Alzheimer's disease (AD) pathology [10]. Rather, we found an association with cerebral infarcts. Other evidence from ROSMP continue to point to potential common mechanisms. For example, we found that brain insulin signaling was associated with AD pathology [11]. We also found interactions between *GSKβ* polymorphisms associated with β-amyloid deposition [12].

The current approaches to identifying several shared pathophysiology processes between Alzheimer's dementia and T2DM have several limitations. Firstly, the most previous works have focused on identifying biological pathways underlying AD and T2MD. Few attempts to discover the role of dysregulated SNPs, gene expressions and methylations have been carried out. Secondly, the conventional evidences for linking AD and T2MD purely depend on the statistical association [13]. Numerous association studies strongly demonstrate that association analysis lacks power to discover the mechanisms of the diseases for the two major reasons. The first reason is that association and causation are different concepts. Association is to characterize the trend pattern between two variables, while causation between two variables is defined as independence between the distribution of cause and conditional distribution of the effect, given cause. There are three scenarios: (1) presence of both association and causation between two variables, (2) presence of association, while absence of causation and (3) presence of causation, while lack of association in causal analysis. If causation loci were searched only from association loci, many causation loci might be missed. The second reason is that the widespread networks that are constructed in integrated omic analysis are undirected graphs. Using undirected graphs, we are unable to infer direct cause-effect relations and hence cannot discover chain of causal mechanism from genetic variation to diseases via gene expressions, epigenetic variation, physiological and phenotype variations. Causal inference coupled with multiple omics, imaging, physiological and phenotypic data is an essential component for the discovery of disease mechanisms.

It is time to develop a new generation of genetic analysis for shifting the current paradigm of genetic analysis from association analysis to deep causal inference and from genetic analysis alone to integrated causal genomic, epigenomic, and phenotypic data analysis for unraveling the mechanic link between AD and T2DM. To make the shift feasible, we need (1) to develop novel causal inference methods for genetic studies of AD and T2DM; (2) to develop unified frameworks for systematic casual analysis of integrated genomic, epigenomic, and clinical phenotype data and to infer multilevel omics causal networks for the discovery of common paths from genetic variants to AD and T2DM via methylations, gene expressions and multiple phenotypes.  The real dada set  ROSMAP [14, 15] will be used to valid the multilevel omics causal networks as a general framework for identifying shared causal paths between AD and T2DM and demonstrates that the proposed methods are capable of identifying the shared pathologic paths between AD and T2DM. A program for implementing the algorithm for construction of multilevel causal networks can be downloaded from our website https://sph.uth.edu/research/centers/hgc/xiong/software.htm.

## Methods

### *ROSMAP Data*

The data came from two longitudinal cohort studies of older persons, ROS that started in 1994 and enrolled Catholic nuns, priests, and brothers from more than 40 communities across United States, and MAP that started in 1997 and enrolled participants with diverse backgrounds and socioeconomic status from continuous care retirement communities throughout northeastern Illinois, as well as from individual homes across the Chicago metropolitan area [34]. These two studies are managed by the same team of investigators. Structured, quantitative neuropathological examinations are performed at a single site. Therefore, the data can be combined in the analysis. Multi-layered omics datasets are generated from biospecimens donated by ROS and MAP participants, including genotypes, DNA methylation profiles and RNA-seq. The genotype data were generated by Affymetrix or the Illumina Omniquad express gene chips and were imputed using the 1000 Genomes Project data as reference. DNA methylation profiles were measured using the Illumina Infinium HumanMethylation450 beadset. RNA-seq data were generated using the Illumina HiSeq with 101bp paired-end reads. Multiple phenotypes including clinical diagnosis, cognitive function, measures of lifestyle, behavior, and activity, chronic medical conditions and risk factors were measured. A total of 432 individuals who simultaneously had genotype, RNA-seq, DNA methylation and some phenotypes were included in analysis. In the analysis, we considered 19 phenotypes and environments, two diseases (AD, T2DM), 299 pathways with RNA-Seq in KEGG pathway database, 20,242 methylation genes with 364,661 CpG sites, and 51, 060 genotyped genes with 5,711,541 SNPs ( 4,283,876 common snp, 1,427,665 rare snp).

***Genome-wide Causation Studies***

Unlike GWAS where we test the association of each variant across the genome with the disease, genome-wide causation studies (GWCS) is to test the causation of each variant across the genome to the disease. The additive noise models (ANMs) with discrete variables will be used for GWCS [27-30]. The procedures that use the ANMs for GWCS are summarized as follows [19].

*Procedures for Causal Genetic Analysis Using ANM:*

1. Fit the following nonlinear integer regression to the data.

   $Y = f(X) + N_Y.$

   Calculate the residuals $\widehat{N}_Y = Y - \hat{f}(X)$.

2. Fit the following nonlinear integer regression to the data.

   $X = g(Y) + N_X.$

   Calculate the residuals $\widehat{N}_X = X - \hat{g}(Y)$ .

3. Test for independence.

   The contingence table and Fisher's exact test can be used to test independence. Let the statistic for testing the independence between $\widehat{N}_Y$ and $X$ as $\Delta_{X\to Y}$ and the statistic for testing the independence between $\widehat{N}_X$ and $Y$ as $\Delta_{Y\to X}$.

The null hypothesis for testing the causation of the variant is

$H_0$ : no causation between variables $X$ and $Y$ .

The statistic for testing the causation between two $X$ and $Y$ is defined as

$$T_C = | \Delta_{X\to Y} - \Delta_{Y\to X} |.$$

When $T_C$ is large, the causation between genetic variant $X$ and disease status $Y$ exists. When $T_C \approx 0$, this indicates that no causal decision can be made. Since the distribution of the test statistic $T_C$ is difficult to calculate, P-value for testing the causation of the variant $X$ can be calculated by permutations.

To improve the performance of causation analysis of rare variants, we first calculate the functional principle component score (FPCS)of the rare variants within a gene [30] to summarize information of all rare variants within the gene . Then, the continuous FPCS are discreterized. Finally, the ANMs with discrete variables can be used to test causation of discreterized FPCS with the disease.

***Structural Equations for Construction of Causal Networks***

Directed graphical models and structural equations can be used as a tool to model the complex causal structures among variables [30]. A graphical model consists of nodes and edges. The nodes represent variables and edges represent the dependence structures among variables. A directed graphic model is defined as the graph in which all the inter-node connections have a direction visually denoted by an arrowhead. Directed acyclic graphics (DAGs) are defined as directed graphics with no cycles. In other words, we can never start at a node $X$, travel edges in the directions of the arrows and get back to the node $X$. A DAG with nodes encodes conditional dependence structure of the variables $Y_1,...,Y_n$. We define the parents of a node as the nodes pointing directly to it. The concept of parents provides an easy way to read off conditional independence from DAGs.

Traditional regressions describe one-way or unidirectional relationships among variables in which the variables on the left sides of the equations are dependent variables and the variables on the right sides of the equations are explanatory variables or independent variables. The explanatory variables are used to predict the outcomes of the dependent variables. However, in many cases, there are two ways, or simultaneous relationships between the variables. Variables in some equations are response variables, but will be predictors in other equations. The variables in equations may influence each other. It is difficult to distinguish dependent variables and explanatory variables. The structural equation models (SEMs) are a powerful mathematic tool to describe such data generating mechanism and infer causal relationships among the variables.

The SEMs classify variables into two class variables: endogenous and exogenous variables. The jointly dependent variables that are determined in the model are called endogenous variables. The explanatory variables that are determined outside the model or predetermined are called exogenous variables. In the genotype-phenotype networks, the phenotype variables such as BMI, cognitive function, working memory, are endogenous variables, age, sex, race, environments and genotypes are exogenous variables. In the genotype-expression networks, the gene expressions are endogenous variables and genotypes are exogenous variables. In the methylation-expression networks, gene expressions are endogenous variables and methylations are exogenous variables.

We consider $M$ endogenous variables. Assume that $n$ individuals are sampled. We denote the $n$ observations on the $M$ endogenous variables by the matrix $Y = [y_1, y_2, ..., y_M]$, where $y_i = [y_{1i}, ..., y_{ni}]^T$ is a vector of collecting $n$ observation of the endogenous variable $i$. Exogenous variables are denoted by $X = [x_1, \ldots, x_K]$ where $x_i = [x_{1i}, \ldots, x_{ni}]^T$ .Similarly, random errors are denoted by $E = [e_1, ..., e_M]$, where we assume $E[e_i] = 0$ and $E[e_i e_i^T] = \sigma_i^2 I_n$ for $i = 1, ..., M$. The linear structural equations for modeling relationships among variables can be written as:

$$\begin{aligned} &y_1\gamma_{11}+y_2\gamma_{21}+\ldots+y_M\gamma_{M1}+x_1\beta_{11}+x_2\beta_{21}+\ldots+x_K\beta_{K1}+e_1=0 \\ &\vdots \qquad\qquad\qquad\qquad\qquad\qquad\qquad \vdots \\ &y_1\gamma_{1M}+y_2\gamma_{2M}+\ldots+y_M\gamma_{MM}+x_1\beta_{1M}+x_2\beta_{2M}+\ldots+x_K\beta_{KM}+e_M=0 \end{aligned} \qquad (1)$$

where the $\gamma$ 's and $\beta$ 's are the structural parameters of the system that are unknown. Variables in the SEMs can be classified into two basic types of variables: observed variables that can be measured and the residual error variables that cannot be measured and represent all other unmodeled causes of the variables. Most observed variables are random. Some observed variables may be nonrandom or control variables (e. g. genotypes, drug dosages) whose values remain the same in repeated random sampling or might be manipulated by the experimenter. The observed variables will be further classified into exogenous variables, which lie outside the model, and endogenous variables, whose values are determined through joint interaction with other variables within the system. All nonrandom variables can be viewed as exogenous variables. The terms exogenous and endogenous are model specific. It may be that an exogenous variable in one model is endogenous in another.

Traditionally, we often select one endogenous variable to appear on the left-hand side of the equation. Specifically, the i-th equation is

$$y_i = y_1\gamma_{1i} + \cdots + y_{i-1}\gamma_{i-1i} + y_{i+1}\gamma_{i+1i} + \cdots + y_M\gamma_{Mi} + x_1\beta_{1i} + \cdots + x_K\beta_{Ki} + e_i, \qquad (2)$$

where $\gamma_{ji}$ is a path coefficient that measures the strength of the causal relationship from $y_j$ to $y_i$, $\beta_{ki}$ is a path coefficient from the exogenous variable to the endogenous variable which measure the causal effect of the exogenous variable $x_k$ on the endogenous variable $y_i$. The coefficients $\gamma_{ji}=0$ and $\beta_{ki}=0$ imply the zero direct influence of $Y_j$ and $x_k$ on $Y_i$, respectively and are usually omitted from the equation. Therefore, equation (2) is reduced to

$$\begin{aligned} y_i &= Y_{-i}\gamma_i + X_i\beta_i + +e_i \\ &= W_i\Delta_i + e_i \end{aligned} \qquad (3)$$

where $Y_{-i}$ is a vector of the endogenous variables after removing variable $y_i$ , $\gamma_i$ is a vector of the path coefficients associated with $Y_{-i}$ , and

$$W_i = [Y_{-i} \quad X_i], \Delta_i = \begin{bmatrix} \gamma_i \\ \beta_i \end{bmatrix}.$$

Multiplying by the matrix $X^T$ on both sides of equation (3), we obtained

$$X^T y_i = X^T W_i \Delta_i + X^T e_i . \qquad (4)$$

Estimation of the parameters in the structural equations is rather complex. It involves many different estimation methods with varying statistical properties. We used two stage least squares (2SLS) method to estimate the parameters. In general, the causal networks are sparse. Using weighted least square and $l_1$ -norm penalization of equation (4), we can form the following optimization problem to estimate the structure of causal network:

$$\min_{\Delta_i} \quad f(\Delta_i) + \lambda \| \Delta_i \|_1$$
$$\text{where } f(\Delta_i) = (X^T y_i - X^T W_i \Delta_i)^T (X^T X)^{-1} (X^T y_i - X^T W_i \Delta_i). \qquad (5)$$

The alternating direction method of multipliers (ADMM) and proximal methods can be used to estimate the parameters and structure of causal network [30, 67, 68].

***Functional Structural Equation Models for Construction of Gene-based Causal Networks***

The SEMs carry out variant by variant analysis. However, the power of the traditional variant-by-variant analytical tools for construction of causal networks with rare variants as exogenous variables will be limited. Large simulations have shown that combining information across multiple variants in a genomic region of analysis will greatly enhance the power to infer causal networks with rare variants as exogenous variables . To utilize multi-locus genetic information, we propose to use a genomic region or a gene as a unit in construction of causal networks and develop sparse structural functional equation models (SFEMs) for causal network analysis.

We define a genotype function. Let *t* be a genomic position. Define a genotype function $x_i(t)$ of the *i*-th individual as

$$x_i(t) = \begin{cases} 2P_q(\mathrm{t}), & \mathrm{QQ} \\ P_q(\mathrm{t}) - \mathrm{P_Q}(\mathrm{t}), & \mathrm{Qq} \\ -2P_Q(\mathrm{t}), & \mathrm{qq} \end{cases}$$

where $Q$ and $q$ are two alleles of the marker at the genomic position *t*, $P_Q(t)$ and $P_q(t)$ are the frequencies of the alleles $Q$ and $q$, respectively. Suppose that we are interested in $k$ genomic regions or genes [ $a_j, b_j$], denoted as $T_j, j = 1,2,...,k$ . We consider the following functional structural equation models (FSEMs):

$$\begin{aligned} &y_1\gamma_{11} + y_2\gamma_{21} + ... + y_M\gamma_{M1} + \int_{T_1} x_1(t)\beta_{11}(t)dt + ... + \int_{T_k} x_k(t)\beta_{k1}(t)dt + e_1 = 0 \\ &y_1\gamma_{12} + y_2\gamma_{22} + ... + y_M\gamma_{M2} + \int_{T_1} x_1(t)\beta_{12}(t)dt + ... + \int_{T_k} x_k(t)\beta_{k2}(t)dt + e_2 = 0 \\ &\vdots \qquad\qquad\qquad\qquad \vdots \qquad\qquad\qquad\qquad \vdots \\ &\mathrm{y}_1\gamma_{1M} + y_2\gamma_{2M} + ... + y_M\gamma_{MM} + \int_{T_1} x_1(t)\beta_{1M}(t)dt + ... + \int_{T_k} x_k(t)\beta_{kM}(t)dt + e_M = 0 \end{aligned} \qquad (6)$$

where $\beta_{ji}(t), j = 1,...,k, i = 1,...,M$ are genetic effect functions.

Functional principal components (FPCs) are efficient summary statistics. The FPCs simultaneously employs genetic information of the individual variants and correlation information (LD) among all variants. For each genomic region or gene, we use functional principal component analysis to calculate principal component function. Let $N$ be the number of sampled individuals. We expand $x_{nj}(t), n = 1,..., N, j = 1,2,...,k$ in each genomic region in terms of orthogonal principal component functions:

$$x_{nj}(t) = \sum_{l=1}^{L_j} \eta_{njl}\phi_{jl}(t), \ j = 1,..., k,$$

where $\phi_{jl}(t), j=1,...,k, l=1,...,L_j$ are the $l$ -th principal component function in the $j$ -th genomic region or gene and $\eta_{njl}$ are the functional principal component scores of the $n$ -th individual. Using the functional principal component expansion of $x_{nj}(t)$ , we can transform the FSEMs (6) into the traditional multivariate SEMs (1).

***Integer Programming for Causal Network learning***

Given the dataset, learning causal networks is the task of finding network structures that best fits the data [22]. We used "score and search" methods to learn causal networks via maximizing the score metrics that characterize the causal networks. The "score and search" algorithms consist of two parts: (1) formulate objective function (global score for the whole network) using the score function for each node and (2) search algorithm.

We collected all nodes with directed edges in the causal network into a DAG, denoted as $G=(V,E)$ . The score (objective function) for the DAG $G$ was defined as

$Score\ (G) = \sum_{j \in V} Score_j\ (G),$

where $\text{Score}_j(G)$ was a score for the node $j$ in the network. The $\text{Score}_j(G)$ was calculated as $f(\Delta_j)$ via solving the optimization problem (5). Therefore, the total score can be decomposed into a sum of score for all nodes in the DAG. In addition, the $\text{Score}_j(G)$ is entirely determined by the parent set of the node $j$ in $G$ . A DAG can be encoded by the set $W=\{W_1,...,W_p\}$ of parent variables for all nodes $V$ in the graph $G$ . We use $C(j,W_j)$ to denote a score function for the pair of node $j$ and its parent set $W_j$ . Therefore, the total score for the DAG $G$ was given by $C(D)=\sum_{i \in V} C(v,W_v)$ .

The learning task is to find a DAG that optimizes the global score $C(D)$ over all possible DAGs $D$ or parent sets [22]:

$$\min_{D} \sum_{i\in V, W_v \in D} C(v, W_v).$$

Integer linear programming (ILP) was used as a search algorithm [22]. A DAG learning was formulated as the ILP as follows. We define a variable $x(W_v -> v)$ to indicate the presence or absence of the parent set $W_v$ in the DAG. In other words, $x(W_v \to v) = 1$ if and only if it is the parent set for the node $v$. The parent set $W_v$ can be an empty set. The objective function for the ILP formulation of a DAG learning can be defined as

$$\sum_{v=1}^{p} \sum_{j_v=1}^{J_v} C(v, W_{j_v}) x(W_{j_v} \to v). \tag{7}$$

The goal was to find a candidate parent set $W_v$ for each node $v$ by optimizing the objective function in (7). It is clear that every DAG can be encoded by a zero-one indicator variable. However, any set of zero-one numbers may not encode a DAG. A set of linear constraints must be posted to make the set of indicator variables to represent a DAG. Without constraints all indicator variables for the parent sets will be equal to either zero or one. These solutions will not form a DAG. The constraints need to be imposed to ensure that the solutions encode a DAG. This constraint that is referred to as convexity constraint, can be expressed as

$$\sum_{i_j=1}^{I_j} x(W_{i_j} \to j) = 1, j = 1, \ldots, p. \tag{8}$$

The convexity constraints (8) can define a directed graph. However, the generated directed graph may have cycles. To eliminate a cycle, we need to impose the following constraint to ensure that any subset $C$ of the nodes $V$ in a DAG must contain at least one node that has no parent in the subset $C$

$$\forall C \subseteq \sum_{j\in C} \sum_{W: W\cap C=\emptyset} x(W \to j) \geq 1, \tag{9}$$

which is referred to as cluster-based constraints. Our goal is to find a candidate parent set $W_j$ for each node $j$ by optimizing objective function (7) subject to the constraints (8) and (9).

The branch and bound method is a popular algorithm ensured to find an optimal solution to the 0-1 ILP problem [22] . Let the LP solution represent "solution of the current linear relaxation". The basic idea of the branch and bound method is to successively divide the ILP problem into smaller problems that are easy to solve and reduce the search space. Briefly, the branch and bound algorithm is summarized as follows. Step 1: Let $\hat{x}$ be the LP solution. Step 2: if there are, valid constraints not satisfied by $\hat{x}$ add them and go to Step 1; otherwise if the solution $\hat{x}$ is an integer then stop, the current problem is solved; otherwise branch on a variable with a non-integer part in $\hat{x}$ to generate two new sub-IP problems. We then again use branch and bound algorithms to solve two sub-ILP problems [22].

***Multilevel Causal Networks***

Multilevel causal omic networks integrated genotype subnetworks, methylation subnetworks, gene expression subnetworks, the intermediate phenotype subnetworks and multiple disease subnetworks into a single connected multilevel genotype-disease networks to reveal the deep causal chain of mechanisms underlying the diseases [30]. ILP was extended from a single causal network estimation to joint multiple causal network estimations to integrate genomic, epigenomic and phenotype data.

For the convenience of discussion, consider $M$ gene expression variables $Y_1,...,Y_M$ , $Q$ methylation variables $Z_1,...,Z_Q$ , and $K$ genotype variables $X_1,...,X_K$ . Let $pa_D(d)$ be the parent set of the node $d$ including gene expression, methylation and genotype variables. Consider three types of SEMs. First, we consider a general SEM model for the gene expression:

$$Y_d = \sum_{i \in pa_D(d)} f_{di}(Y_i) + \sum_{q \in pa_D(d)} f_{dq}(Z_q) + \sum_{j \in pa_D(d)} f_{dj}(X_j) + \varepsilon_d \ , \ d = 1, \ldots, M, \quad (10)$$

and

$$Z_q = \sum_{l \in pa_Q(q)} f_{ql}(Z_l) + \sum_{m \in pa_Q(q)} f_{qm}(X_m) + \varepsilon_q \,, q = 1, \ldots, Q, \quad (11)$$

where $f_d$ and $f_q$ are linear functions from $R^{|pa_D|} \to R$ and $R^{|pa_Q|} \to R$, respectively, and the errors $\varepsilon_d$ and $\varepsilon_q$ are independent , following distributions $P_{\varepsilon_d}$ and $P_{\varepsilon_q}$, respectively. Equation (10) define a causal network that connects gene expressions, methylations and genotypes. Equation (11) define a causal network that connects methylations and genotypes.

*Integer Programming as a General Framework for Joint Estimation of Multiple Causal Networks*

We collected multiple types of data: genotype, gene expression, methylation, and phenotype and disease data. We wanted to estimate multiple causal networks with different types of data.

The scores of the nodes $Y_d$ and $Z_q$ were, respectively, given by

$$C(Y_d, W_{di}) = Y_d^T (I - D_Y^i ((D_Y^i)^T D_Y^i)^{-1} (D_Y^i)^T) Y_d \quad (12)$$

and

$$C(Z_q, W_{ql}) = Z_q^T (I - D_Z^l ((D_Z^l)^T D_Z^l)^{-1} (D_Z^l)^T) Z_q \,, \quad (13)$$

where matrices $D_Y^i$ and $D_Z^l$ corresponded to the parent sets $W_{di}$ and $W_{ql}$.

Let $V_E$ be the set of nodes in the gene expression network and $V_M$ be the set of nodes in the methylation network. Let $C_E$ be a subset of nodes in $V_E$ and $C_M$ be a subset of nodes in $V_M$. A joint expression and methylation causal network can be formulated as the following ILP:

$$\text{Min} \quad \sum_{d=1}^{M} \sum_{i \in pa_D(d)} C(d, W_{di}) \chi(W_{di} \to d) + \sum_{q=1}^{Q} \sum_{l \in pa_{Q(q)}} C(q, W_{ql}) \chi(W_{lq} \to q)$$

$$\begin{aligned} \text{s.t.} \quad & \sum_{\mathrm{i} \in \mathrm{pa_D}(d)} \chi(W_{di} \to d) = 1, d = 1, \ldots, M, \\ & \sum_{\mathrm{l} \in \mathrm{pa_Q}(q)} \chi(W_{ql} \to q) = 1, \; q = 1, \ldots, Q, \\ & \forall C_E \subseteq V_E : \sum_{\mathrm{d} \in \mathrm{C_E}} \sum_{\mathrm{W_d}: W_d \cap C_E = \phi} \chi(W_d \to d) \geq 1, \\ & \forall C_M \subseteq V_M : \sum_{q \in C_M} \sum_{W_q: W_q \cap C_M = \phi} \chi(W_q \to q) \geq 1. \end{aligned} \quad (14)$$

Using branch and bound and other methods for solving the ILP, we can solve the ILP problem (14) to obtain the best joint causal genotype-methylation-expression and genotype-methylation network fitting the data.

*Summary Statistics for Representation of Groups of Gene Expressions*

Generalized low rank models were used to segment (cluster) the data. Principal component analysis (PCA) was used to reduce data dimensions. The PCs were used to summarize the gene expression data in pathways and clusters [69].

## Results

### Shared genetic loci underlying AD and T2DM.

AD and T2DM result from the interplay of DNA sequence variation and nongenetic factors acting through molecular networks [16-18]. Their etiology is complex with multiple steps between genes and phenotypes. Neither traditional GWAS, nor classical multi-omics analysis can identify the causal paths from genetic variants to diseases because not all these analyses can identify directed paths from genetic variants to diseases through environments, methylations, gene expressions, and phenotypes. To overcome these limitations, we develop a novel general framework for identifying all possible causal paths from genetic variants to diseases. The framework consists of three steps. The first step is to perform genome-wide causation studies (GWCS) where we test causation of each SNP across the genome to the disease. The additive noise model (ANM) with discrete variants will be used to test for causation [19] (Methods). We focused on the rare variants in the paper. The second step is to use integer programming (IP) and various modern causal models [20-22] (Methods) for inferring multilevel genome-wide omic causal networks that integrate genotype subnetworks, environmental subnetworks, methylation subnetworks, gene regulatory subnetworks, intermediate phenotype subnetworks and multiple disease subnetworks into a single connected multilevel genotype-disease network as shown in Figure 1. The third step is to augment graph theoretical approaches with approximations for developing efficient search algorithms that discover all possible paths starting from the genetic variant node directed to the disease node, including classical Depth First Search (DFS) and Breadth First Search (BFS) algorithms [23-26].

There are two ways to identify shared dysfunctional genes (SNPs) between AD and T2DM. One way is to use ANM with discrete variables and functional data analysis to conduct genome-wide causation analysis [27-30] for unravelling the direct connections between gene nodes and disease nodes to identify the shared dysfunctional genes between AD and T2DM.

Another way is to search the paths from the gene nodes to AD and T2DM in multilevel causal omics networks.

Association and causation are different concepts. Association between two variables is often characterized by dependence between two variables. Causation is a connection of phenomena where one variable acts or intervenes on another variables and leads to its changes. Therefore, the key component of causation is the generation and determination of values of one variable by another. The mechanism of causation is related to the transference of matter, motion and information. Causation is universe. It is a part of universe connection. It is well known that nature consists of autonomous and independent causal generating process modules. These modules will not influence each other [29, 31]. In other words, while output of one module may inform or influence input of another module, the events between modules are independent. In the probabilistic language, mechanism is often represented by conditional distribution. Independent mechanism states that "the conditional distribution of each variable given its causes (i.e., its mechanism) does not inform or influence the other conditional distributions" [29]. In GWCS, we only consider two variables. In this case, independence of cause and mechanism (ICM) indicates that the conditional distribution of the effect given its cause is independent of distribution of cause. Consider the genetic analysis of alleles $(A)$ with a disease allele $A$ a normal allele $a$ and with the disease$(D)$ (disease $D$ and normal $d$ ). The joint density function $P(a, d)$ can be decomposed into

$$P(A, D) = P(A)P(D|A)$$
$$= P(D)P(A|D).$$

In the association analysis, we assess whether $A$ is independent of $D$ or not. The relationship between $A$ and $D$ is symmetric. However, in causal analysis, causations $A \rightarrow D$ and $D \rightarrow A$ are different. They are asymmetric. Assessing causation is to consider the effect of intervention. Causation $A \rightarrow D$ indicates that the effect of $A$ is to give rise to disease. However, disease status $D$ will not generate allele $A$. Suppose that locus $A$ is disease locus and $A \rightarrow D$. If we change the allele $a$ to allele $A$, then we assume that biological mechanism $P(D|A)$ responsible for giving rise to disease. This would hold true independent of the distribution (frequencies)_of allele $A$. If the locus A is disease locus, we can find that the distributions (frequencies) of allele $A$ in two different populations are different, but the mechanism $P(D|A)$ would apply in two population. The conditional probability $P(D|A)$ can also be viewed as penetrance of the allele. The marginal distribution $P(A)$ and conditional distribution $P(D|A)$ contain no information about each other. Both continuous and discrete ANMs satisfy the ICM and will be used for GWCS., The proposed method for genome-wide causation analysis and inferring multilevel causal genotype-methylation-expression-phenotype-disease network was applied to the ROSMAP dataset [34] with 432 individuals, 19 phenotypes and environments, two diseases (AD, T2DM), 299 pathways with RNA-Seq in KEGG pathway database, 20,242 methylation genes with 364,661

CpG sites, and 51, 060 genotyped genes with 5,711,541 SNPs ( 4,283,876 common snp, 1,427,665 rare snp ) (imputed by 1000 Gnome Data). The inferred genotype-expression-methylation-phenotype-disease network consisted of 2,814 nodes and 22,184 edges where the edges were presented in the network if the path coefficients were significantly from zero with P-values $< 0.05$.

Table 1. The number of genes connected to AD and T2DM.

| | | To T2DM | | | |
|---|---|---|---|---|---|
| | | Directly Connected | Indirectly Connected | Both Directly and Indirectly Connected | Not Connected |
| To AD | Directly Connected | 5 | | | 13 |
| | Indirectly Connected | | 682 | 13 | |
| | Both Directly and Indirectly Connected | | 20 | 8 | |
| | Not Connected | 17 | | | |

There were two ways to connect a gene (or SNP) to AD (T2DM). If a gene (or SNP) showed causation to AD (T2DM) by statistical causal test, then the gene (SNP) was directly connected to AD (T2DM) in the causal network. Such gene (SNP) was called AD (T2DM) directly connected gene (SNP). We may observe the connection between a gene (SNP) and AD (T2DM) via multiple edges (paths) in the constructed multilevel causal network. Then, the gene (SNP) that was indirectly connected to AD (T2DM) via paths in the multilevel causal network was called AD (T2DM) indirectly connected gene (SNP). The number of AD and T2DM directly connected or indirectly connected genes was summarized in Table 1. The total number of genes connected to both AD and T2DM including directly connected and indirectly connected was 759. The genes that were both directly and indirectly connected to both AD and T2DM were summarized in Table S1. The genes that were indirectly connected to AD and both directly and indirectly connected to T2DM were listed in Table S2. Similarly, the genes that were both directly and indirectly connected to AD and indirectly connected to T2DM were summarized in Table S3.

We also tested causation of 299 pathways in the KEGG pathway database to AD and T2DM (Described in detail in the Methods section). The results were summarized as follows. The number of pathways that were directly connected to both AD and T2DM was 16; the number of pathways that were directly connected to AD and indirectly connected to T2DM was 17; the number of pathways that were directly connected to T2DM and indirectly connected to AD was 18, the number of pathways that were indirectly connected to both AD and T2DM was 114; the number of pathways that were directly connected to AD and not connected to T2DM was 6; the number of pathways that were not connected to AD and directly connected to T2DM was 2.

Then, we investigated shared gene expressions via multilevel causal networks. We summarized the results as follows. The number of expression genes that were directly connected to both AD and T2DM was two genes: GRMD1B, RP1-111D6.3, the number of expression genes that were directly connected to AD, but not directly connected to T2DM was 19 (P-value $< 10^{-4}$, Table S4**)** and the number of expression genes that were directly connected to T2DM, but not directly connected to AD was 7 (P-value $< 10^{-4}$, Table S5). The number of expression genes that were indirectly connected to both AD and T2DM was 725.

Similarly, we can study shared methylation via multilevel causal networks. The number of methylated sites/ genes that were directly connected to AD, but not directly connected to T2DM was 17 (Table S6) and the number of methylated sites/genes that were directly connected to T2DM, but not directly connected to AD was 27 (Table S7). The number of methylated sites/genes that were indirectly connected to both AD and T2DM was 117 (Table S8).

The number of phenotypes that were directly connected to both AD and T2DM was six (Age, CHL, HDL ratio, LDL, Semantic memory and working memory).

**Shared CREBBP, MAPK and PI3K-AKT pathways between AD and T2DM**

Binding of transcription factors to the cyclic Adenosine Monophosphate (cAMP) response element (CRE) regulates the activity of RNA polymerase. cAMP Response Element binding protein (CREB) is a cellular transcription factor that binds the CRE [32]. CREB-binding protein (CREBBP) and CREB together mediate the conversion of short-term memory to long-term memory and alternate the activity of the β-amyloid (Aβ) peptide, which in turn regulates hippocampal-dependent synaptic plasticity [33, 34]. Cognitive function such as working memory is involved in insulin signaling dysfunction and blood glucose levels. It was reported that working memory is linked with T2MD [35-37].

To assess whether CREBBP is a common genetic factor of AD and T2DM, and how CREBBP mediates the development of AD and T2DM, we searched the all possible paths from gene CREBBP to AD and T2DM in the inferred multilevel causal network. The results were shown in Figure 2. Figure 2A plotted the path from CREBBP to AD and T2DM via MAPK and PI3K-AKT signaling pathways. The genes in the MAPK and PI3K-AKT signaling pathways, CREBBP, episodic memory, MMSE, AD and T2DM were then used to further infer causal networks using SEMs and IP. The inferred causal network was shown in Figure 2B. From Figure 2B we observed a path from CREBBP to AD and T2DM via gene connections: $CREBBP \rightarrow CBL \rightarrow MAP2K4 \rightarrow MAPK8 \rightarrow MAPK1 \rightarrow PIK3CA$. MAPK and PI3K-AKT pathways play critical roles in memory.

**Shared TTC3, FoxO, MAPK, and PI3K-AKT Pathways between AD and T2DM**

Next we presented an example to illustrate shared causal paths that started a gene directly connected to AD and indirectly connected to T2DM. The tetratricopeptide repeat domain 3 (TTC3) gene was an AD causing gene (P-value for causation of AD $< 0.0001$), but not directly connected to T2DM (P-value for causation of T2DM =0.47). TTC3 is associated with differentiation of neurons [38]. It is reported that a rare TTC3 variant is related with AD [39]. The TTC3–RhoA pathway could be a key determinant of the neuronal development, resulting in detrimental effects on the normal differentiation program [40]. Rho regulates the activation of MAPK pathway [41]. The Forkhead box O (FoxO) transcription factors that affect nervous system amyloid (Aβ) production, are implicated in the regulation of cell apoptosis and survival, and accelerate the progression of degenerative disease. FoxO pathway is involved in the PI3K/Akt and mitogen-activated protein kinase (MAPK) pathways in neuronal apoptosis in the brain.

FoxOs also can offer protection in the nervous system, reduce toxic intracellular protein accumulations and have potential to limit Aβ toxicity [42, 43, 34]. Akt-FoxO that suppresses TLR4 signaling in Human Leukocytes is implicated in the development of T2DM [44]. There are increasing evidences that PI3K/AKT pathway are implicated in the development of T2DM [45, 46].

Again, we used the DFS algorithm to search the causal paths from multilevel causal networks. The causal paths from TTC3 to AD and T2DM were shown in Figure 3. The paths from MAPK and PI3K-AKT pathway to AD and T2DM were the same as that in Figure 2. The genes in the FoxO, MAPK and PI3K-AKT signaling pathways, TTC3, and episodic memory, MMSE, weight, AD and T2DM were then used to further infer causal networks using SEMs and IP. The structure of the inferred network was shown in Figure 3B. There were a large number of causal paths from *TTC3* to either AD or T2DM. The shared common causal paths were $TTC3 \rightarrow NLK \rightarrow CACNA2D1 \rightarrow CNCNG3 \rightarrow FOXO1 \rightarrow CCNE1 \rightarrow CYCS \rightarrow MAPK1 \rightarrow PIK3CA$ and $TTC3 \rightarrow NLK \rightarrow PLK2 \rightarrow MAPK8 \rightarrow MAPK1 \rightarrow MAPK1 \rightarrow PIK3CA.$

**Shared Morphine Addiction and Neuroactive Ligand Receptor Interaction Pathways**

Morphine addiction has neurotoxic effects and damages to the brain regions that function for learning, memory and emotions [47]. High dose of morphine may increase risk to T2DM [48]. It is also reported that neuroactive ligand receptor interaction pathway is associated with both AD and T2DM [49].

Searching the causal paths from gene *HNF4G* to AD and T2DM via the multilevel causal networks using the DFS algorithm, we found that *HNF4G* was indirectly connected to AD and T2DM. In addition to shared MAPK and PI3K-AKT pathways between AD and T2DM which were discussed in the previous sections, we observed shared two new pathways between AD and T2DM: morphine addiction and neuroactive ligand receptor interaction pathways as shown in Figure S1A. The structure of the inferred network that consisted of shared morphine addiction and neuroactive ligand receptor interaction pathways between AD and T2DM was shown in Figure S1B. There were more than 10 shared causal paths. We observed two shared major causal paths: (1) $HNF4G \rightarrow NLK \rightarrow PLK2 \rightarrow MAPK8 \rightarrow MAPK1 \rightarrow PIK3CA \rightarrow AKT1$ amd (2) $HNF4G \rightarrow NLK \rightarrow GNGT2 \rightarrow PLCB2 \rightarrow PLCB1 \rightarrow ADRB1.$

**Shared Fatty Acid Biosynthesis and Primary Bile Acid Biosynthesis Pathways**

Brain function such as intelligence, memory, behavior and concentration are all influenced by brain nutrition [50]. Omega-3 fatty acids affect the fluidity of brain cell membranes, neurotransmitter synthesis and signal transmission and are implicated in AD [51, 52]. Bile acids are involved in cell signaling and immune function. It acts as potent inhibitors of apoptosis and regulates transcriptional and post-transcriptional events that affect mitochondrial function in neurons [53]. A trend of increased bile acids in AD has been observed [54]. Fatty acid utilization induces insulin resistance [55]. Bile acids are signal molecules and play an important role in regulating metabolism and inflammation. The abnormal bile acids are correlated with changes in insulin secretion, which lead to T2DM [56, 57]. The amyloid precursor protein (APP) is a transmembrane protein. The aggregated amyloid-β (Aβ) peptides are generated by sequential proteolytic processing of the APP. Accumulation of Aβ and the APP play an important role in regulating lipid homeostasis including fatty acids, which finally affect the development of AD [58].

Our data also provided evidence to show that fatty acid biosynthesis and primary bile acid biosynthesis pathways were shared pathways between AD and T2DM. Search the multilevel causal networks from APP to AD and T2DM using the DFS algorithm, we identified the shared causal paths from APP to both AD and T2DM, shown in Figure 4A. There were two shared causal paths between AD and T2DM: $APP \rightarrow$ $neuroactive\ ligand\ receptor\ interaction$ and $APP \rightarrow fatty\ acid\ biosynthesis\ \rightarrow primary\ bile\ acid\ biosynthesis$. Neuroactive ligand receptor interaction pathway was discussed in the previous section.

Next we presented the causal network structure of the shared genes between AD and T2DM in the two shared causal paths in Figure 4B. We observed two major shared paths from APP to AD and T2DM. One path was $APP \rightarrow ACSL4 \rightarrow ACACA \rightarrow NUDT9 \rightarrow CMC1 \rightarrow PTPLAD1 \rightarrow CYP781 \rightarrow CYP46A1 \rightarrow working\ memory\ (or\ CYP781 \rightarrow AMACA \rightarrow working\ memory)$. Another causal path was $APP \rightarrow F2RL3 \rightarrow PIK3R3\ (or\ F2RL3 \rightarrow S1PR3 \rightarrow PIK3R3)$.

To further illustrate the validity of the inferred causal paths, we presented Figure S2 that showed the average levels of expression of the genes in Figure 4 for AD, T2DM and normal individuals. From Figure 4, Figures S2 and S3, we can observed that the genes along the path $APP \rightarrow F2RL3 \rightarrow PIK3R3$ $(or\ F2RL3 \rightarrow S1PR3 \rightarrow PIK3R3)$ of the individuals with AD were over expressed, and the genes along the path $APP \rightarrow ACSL4 \rightarrow ACACA \rightarrow NUDT9 \rightarrow CMC1 \rightarrow PTPLAD1 \rightarrow CYP781 \rightarrow CYP46A1 \rightarrow working\ memory$ $(or\ CYP781 \rightarrow AMACA \rightarrow working\ memory)$ of the individuals with AD were under expressed. Genetic variation in gene *APP* either regulated over expressed genes or regulated under expressed genes. Both of them caused AD. For the individuals with T2DM, the majority of gene expressions along the causal paths from *APP* to T2DM which were regulated by genetic variation in gene *APP* was under expressed.

**Shared Methylated Genes POU3F2, KIF4B and TNSL3, and Dopaminergic Synapse and AMPK Pathways**

In this section, we illustrate how a shared gene regulates three shared gene methylations, which in turn regulate the shared pathways. Emerging evidences indicate that methylation alternations to DNA of the brain are linked to Alzheimer's disease [62, 63]. DNA methylation also plays an important role in the pathogenesis of T2DM [63, 64]. In order to better understand the etiology of AD and T2DM, we jointly investigated the genetic variants, DNA methylation and gene expression profiles, multiple phenotypes, AD and T2DM using causal inference pipelines. We found that gene *POU3F2* regulated methylations of POU3F2, KIF4B and TMSL3. Alternations in methylation of three genes directly caused the development of AD and T2DM. Furthermore, methylation levels of three genes regulated gene expressions in dopaminergic synapse and AMPK pathways, which in turn caused AD and T2DM via CHL/HDL Ratio (Figure 5A). Recent advance revealed that Alterations of the dopaminergic system contributes to memory and reward dysfunction and the dopaminergic system may well be involved in the occurrence of AD [59, 60]. Recent studies also unravel that the brain damage in AD is linked to an over-activation of AMPK, which leads to the loss of the ability of neurons to grow axons and the modification of the tau proteins resulting in tangles of tau [65]. The AMPK functions as a key energy sensor. AMPK signaling elicits insulin-sensitizing effects and may be implicated in stimulating glucose up taking in skeletal muscles, fatty acid oxidation in adipose (and other) tissues [66]. Our results showed that genetic variation in gene *POU3F2* regulated gene expressions in dopaminergic synapse and AMPK pathways via methylations of *POU3F2*, *KIF4B* and *TMSL3*, which in turn influences CHL/HDL Ration, and finally led to AD and T2DM (Figure 5A).

Again, we presented the causal network structure of the shared genes between AD and T2DM in the two shared dopaminergic synapse and AMPK pathways in Figure 5B. There were multiple shared directed paths from POU3F2 to AD and T2DM. A major shared directed path:

$$m{:}POU3F2 \rightarrow m{:}LOC644649 \rightarrow KDM5C \rightarrow PDPK2 \rightarrow XPA \rightarrow MK3R2 \rightarrow ELK1 \rightarrow AD\ (or\ \rightarrow CHL \rightarrow T2DM).$$

## Discussion

This paper addresses several issues for uncovering causal paths shared between AD and T2DM. The first issue is to shift the current paradigm of genetic analysis from association analysis to deep causal inference for uncovering the shared mechanisms between AD and T2DM. The current paradigm for discovering mechanisms of diseases is association analysis. There is increasing recognition that a large proportion of association signals are not causal signals and causal signals may not be association signals. A large number of causal signals cannot be derived from set of association signals. Only searching causal signals from association analysis, a large proportion of causal signals will be missing. Therefore, the ANMs were developed as practical causal inference methods to identify the genetic variants that cause disease.

Second issue is to shift the current paradigm of genetic analysis from genetic analysis alone to integrated causal genomic, epigenomic, transcriptional and phenotypic data analysis for unraveling the mechanisms of AD and T2DM. The widespread existing omics networks that are constructed in integrated multistep analysis of omics are undirected graphs. Using undirected graphs, we are unable to infer direct cause-effect relations among diversified types of variables at multilevel and hence cannot discover chain of causal mechanism from genetic variation to diseases via omics. In this paper, we develop novel statistical methods for multilevel causal omics network construction and provide pipelines for uncovering shared causal paths between AD and T2DM via gene expressions, DNA methylations, environments and multiple phenotypes. The third issue is to develop algorithms that can automatically search the causal paths from genetic variants to diseases. The size of multilevel causal omics network is large. The number of nodes of such networks can reach ten thousands. The number of causal paths is huge. Manually searching causal paths from large causal networks is infeasible. To meet the challenge of searching causal paths from large causal networks, we develop computer representation of large causal networks and algorithms for searching the causal paths.

## Conclusion

The results of application of the proposed pipelines for identifying causal paths to real data analysis of AD and T2DM provided strong evidence to support the link between AD and T2DM and unraveled causal mechanism to explain this link. We identified the shared causal genes, gene expressions, DNA methylations and pathways between AD and T2DM. Some of them can be supported by literature and some of them are new. We identified an extremely large number of shared causal paths from genetic variants to both AD and T2DM via DNA methylation, gene expressions and phenotypes. This deep knowledge that uncovered the large number of causal mechanisms of AD and T2DM had profound implication in prevention and treatments of AD and T2DM. This explained why the drugs that were based on inhibition or activation of limited number of paths often failed simply because these limited number of paths cannot cover all causal paths to the diseases. Finally, the empirical evidence that the AD and T2DM shared a large number of causal genes, gene expressions, methylations and pathways supported hypothesis that AD can be considered as “type 3 diabetes”.

**Acknowledgements**

Mr. Hu and Mr. Jin were supported by National Natural Science Foundation of China (31521003)， Shanghai Municipal Science and Technology Major Project (2017SHZDZX01), and the 111 Project (B13016) from Ministry of Education.

**Author contributions**

Z.H, R. J, P. W, Y.Z developed software and conducted data analysis, M.X designed project and wrote manuscript, L. J, J. Z and J. W designed the project, D. B provided data and wrote manuscript, P. J provided data.

**Additional information**

**Complete financial interests.** The authors declare no competing financial interests.

**Figure**

**Figure 1.** Scheme of multilevel omic networks.

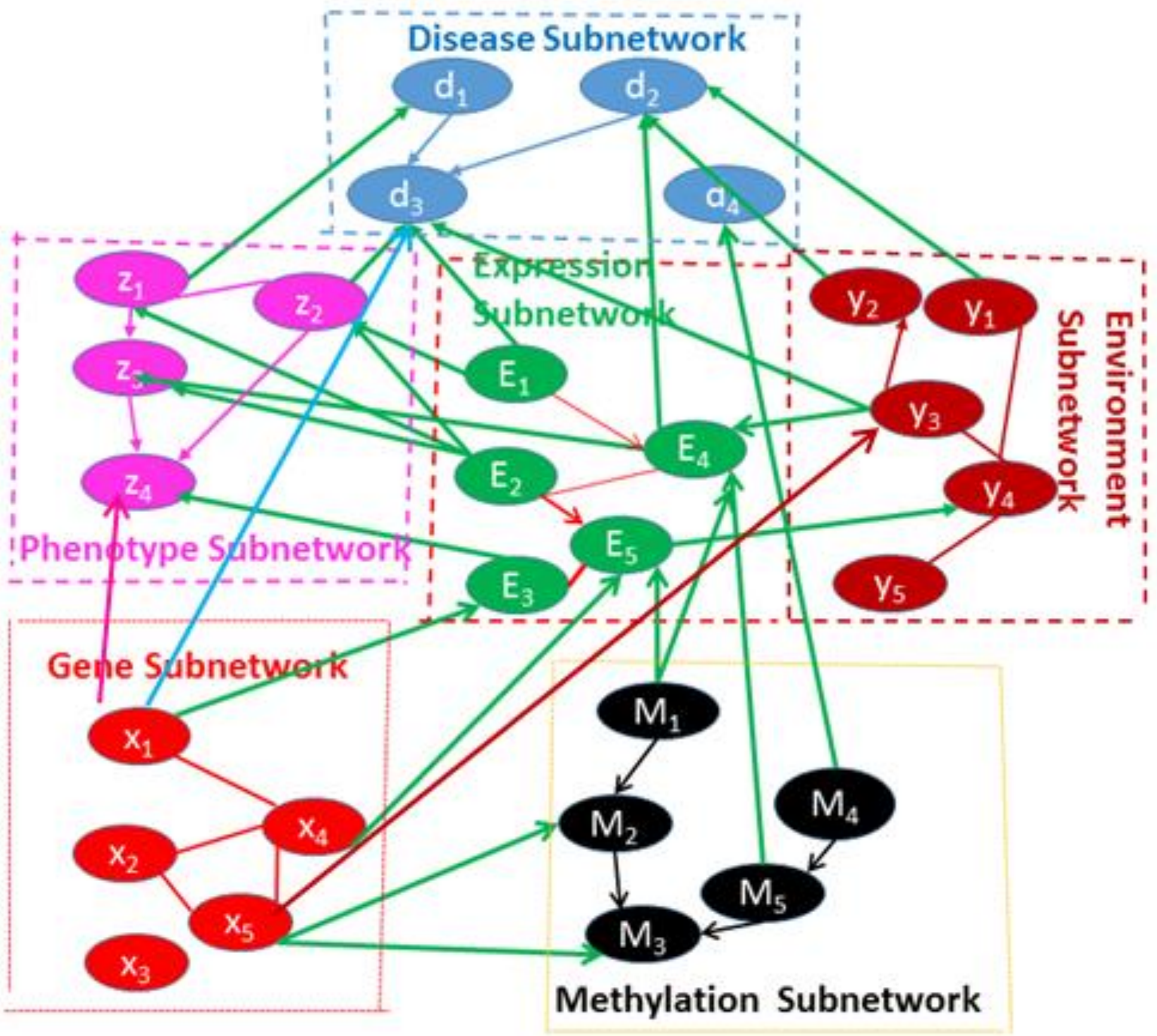

**Figure 2.** (A) Shared CREBBP, MAPK and PI3K-AKT pathways between AD and T2DM; (B) Shared causal subnetwork structure from CREBBP to AD and T2DM.

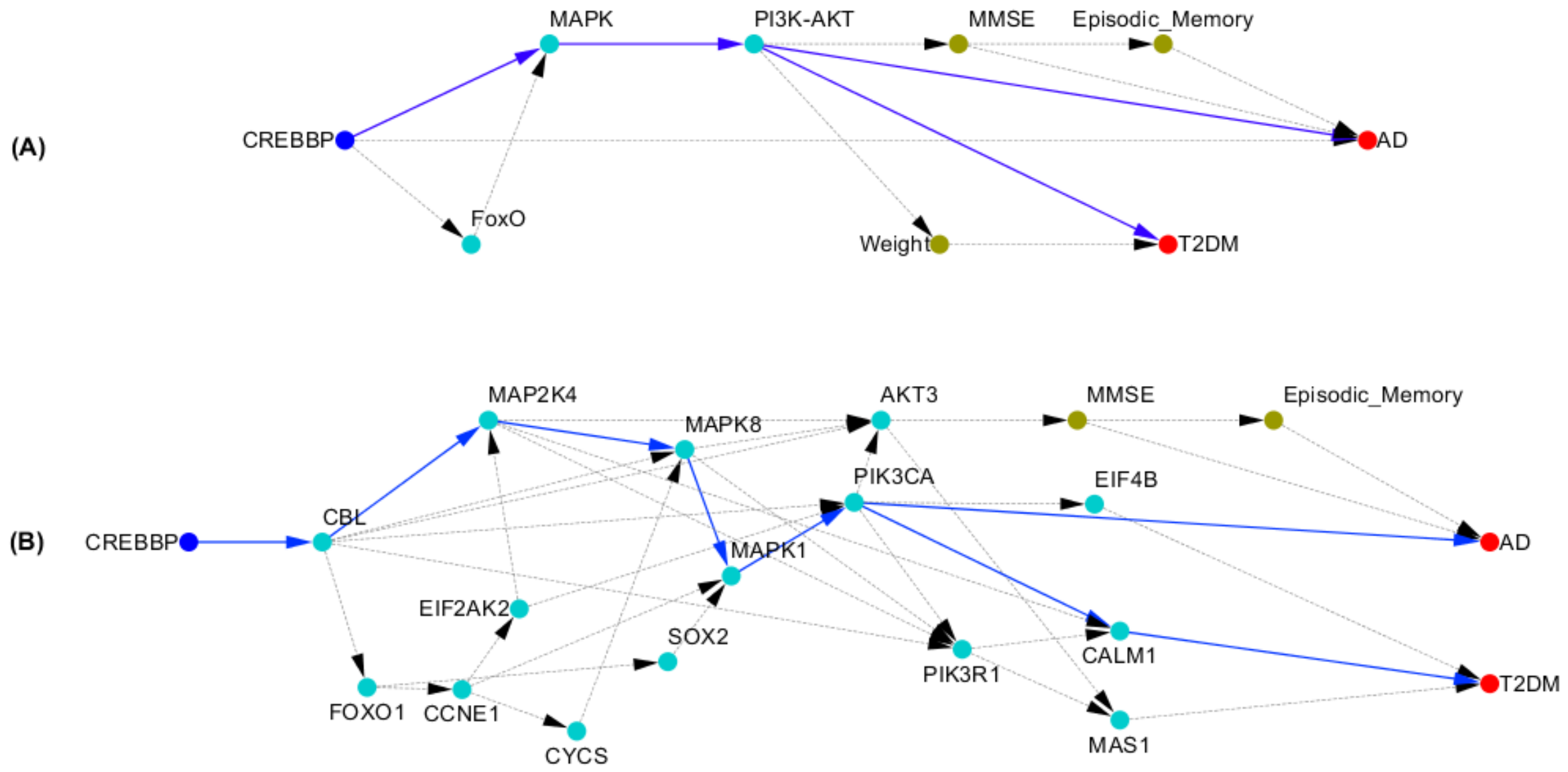

**Figure 3.** (A)Shared TTC3, FoxO, MAPK, and PI3K-AKT Pathways between AD and T2DM; (B) Shared causal subnetwork structure from TTC3 to AD and T2DM.

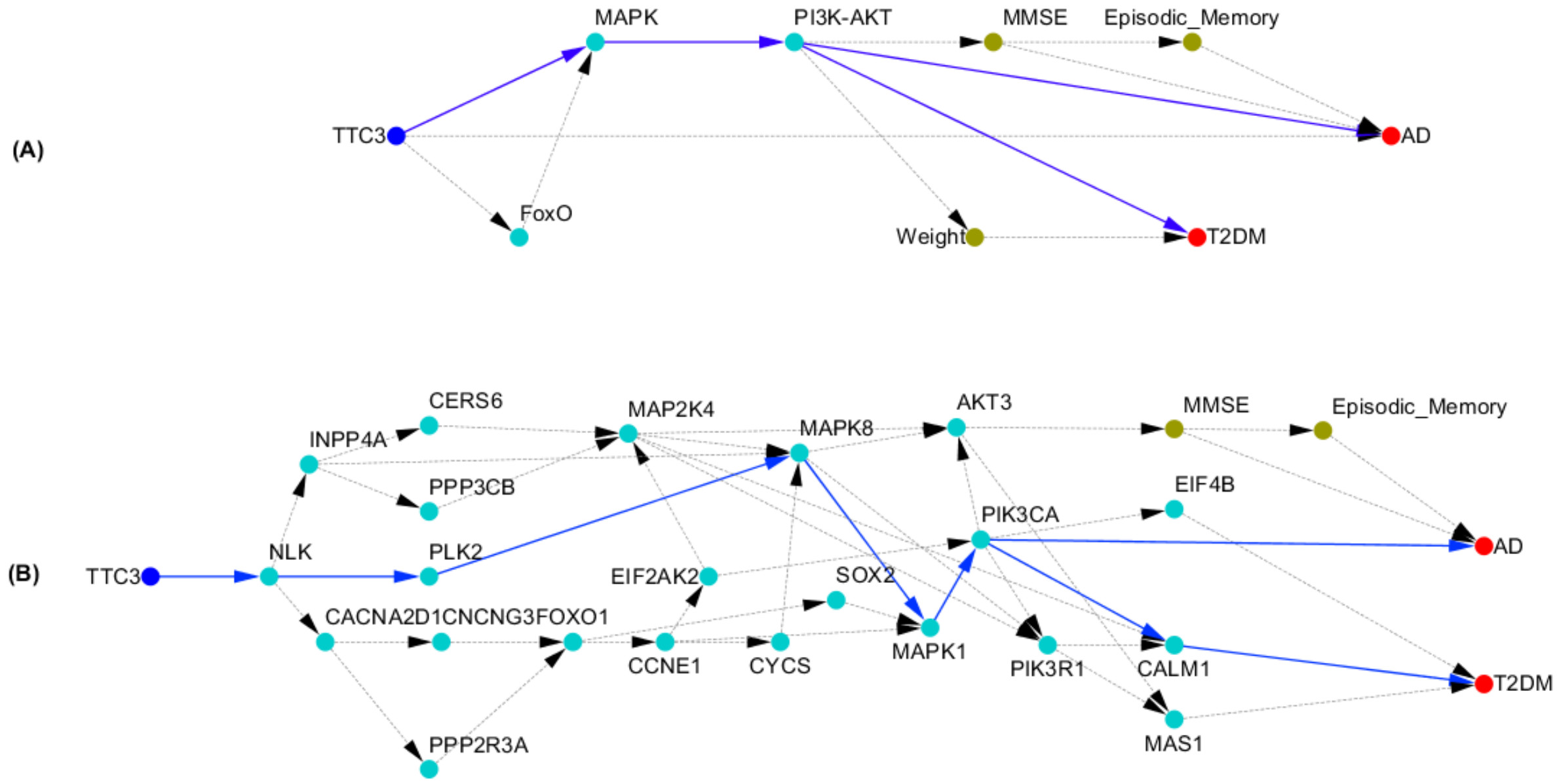

**Figure 4.** (A) Shared APP, Fatty Acid Biosynthesis and Primary Bile Acid Biosynthesis Pathways between AD and T2DM; (B) Shared causal subnetwork structure from APP to AD and T2DM.

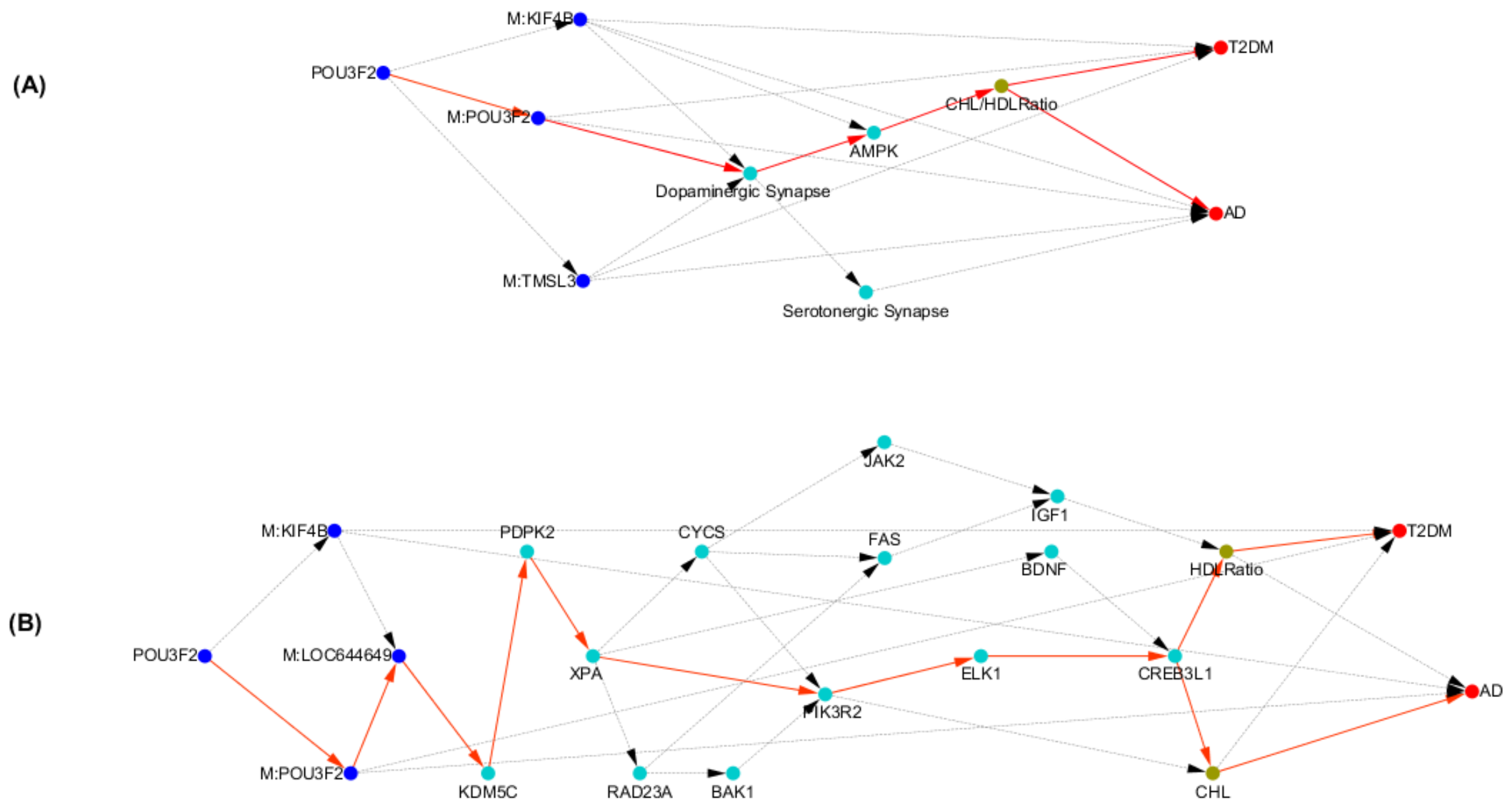

**Figure 5.** (A) Shared Methylated Genes POU3F2, KIF4B and TNSL3, and Dopaminergic Synapse and AMPK Pathways between AD and T2DM; (B) Shared causal subnetwork structure from POU3F2 to AD and T2DM.

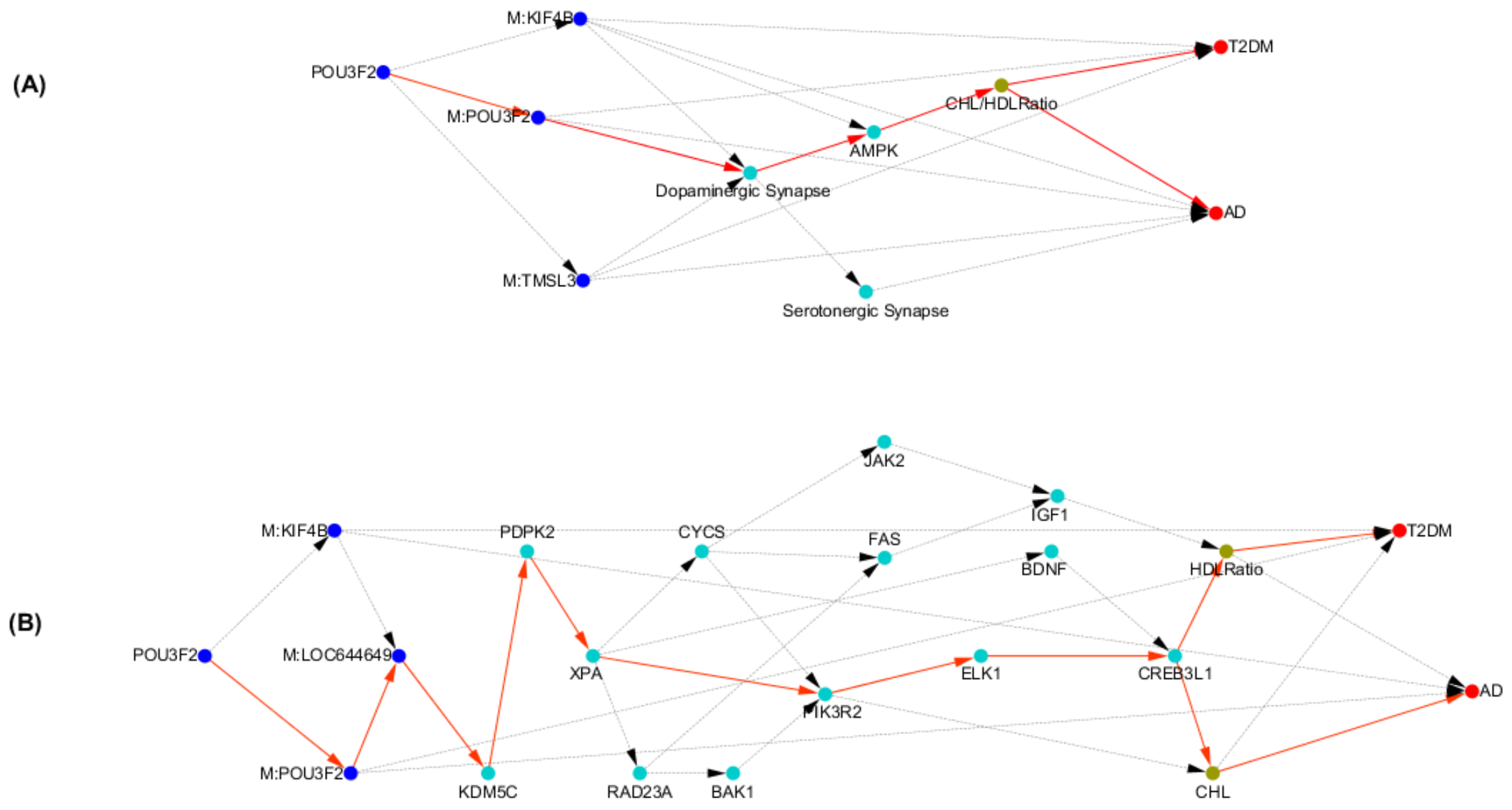

## Supplementary Figure

**Figure S1**. (A) Shared morphine addiction and neuroactive ligand receptor interaction pathways between AD and T2DM; (B) Shared causal subnetwork structure from HNF4G to AD and T2DM.

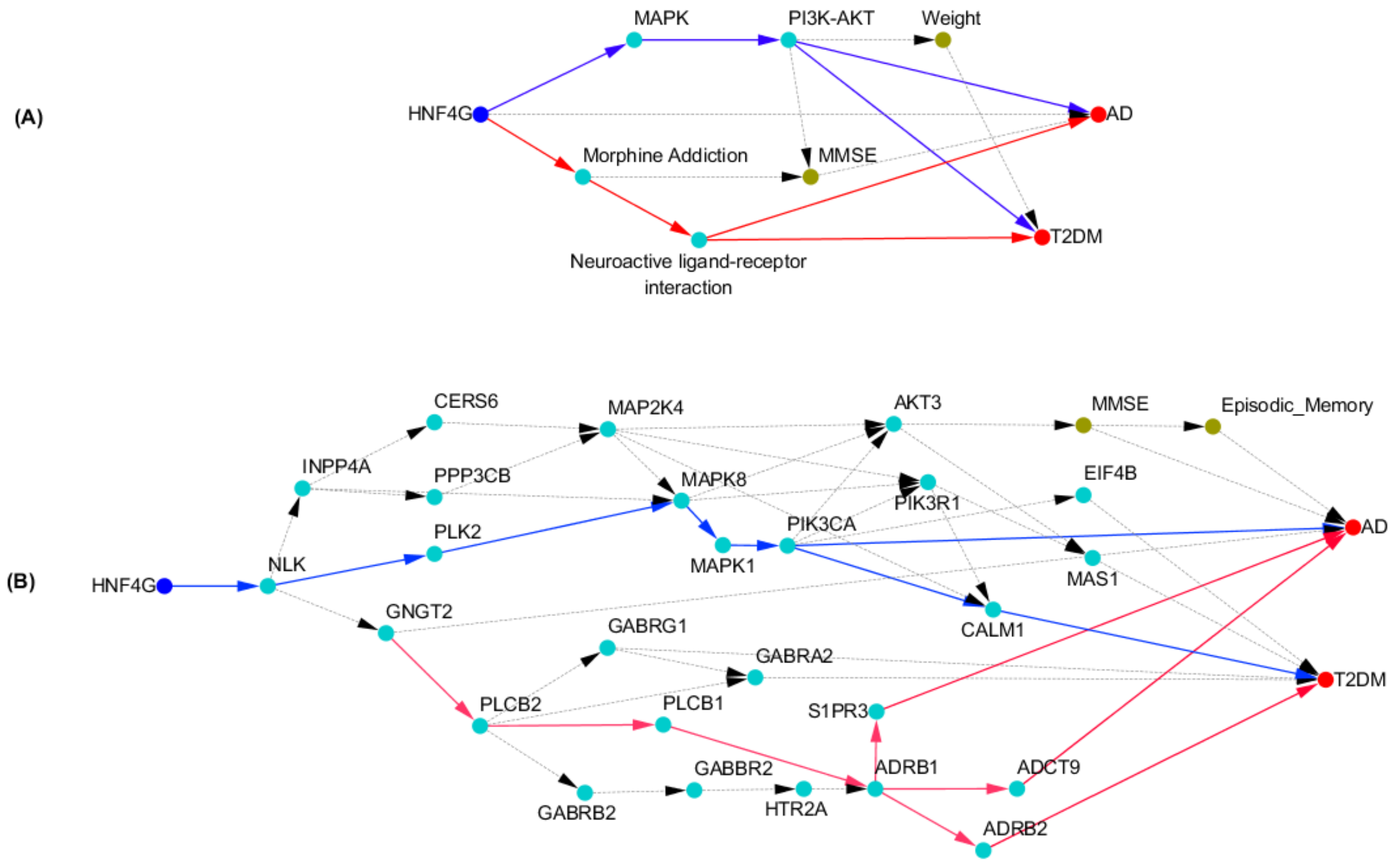

**Figure S2.** Average expression levels of genes in Figure 4 for AD, T2DM and normal individuals where gene expression levels were normalized.

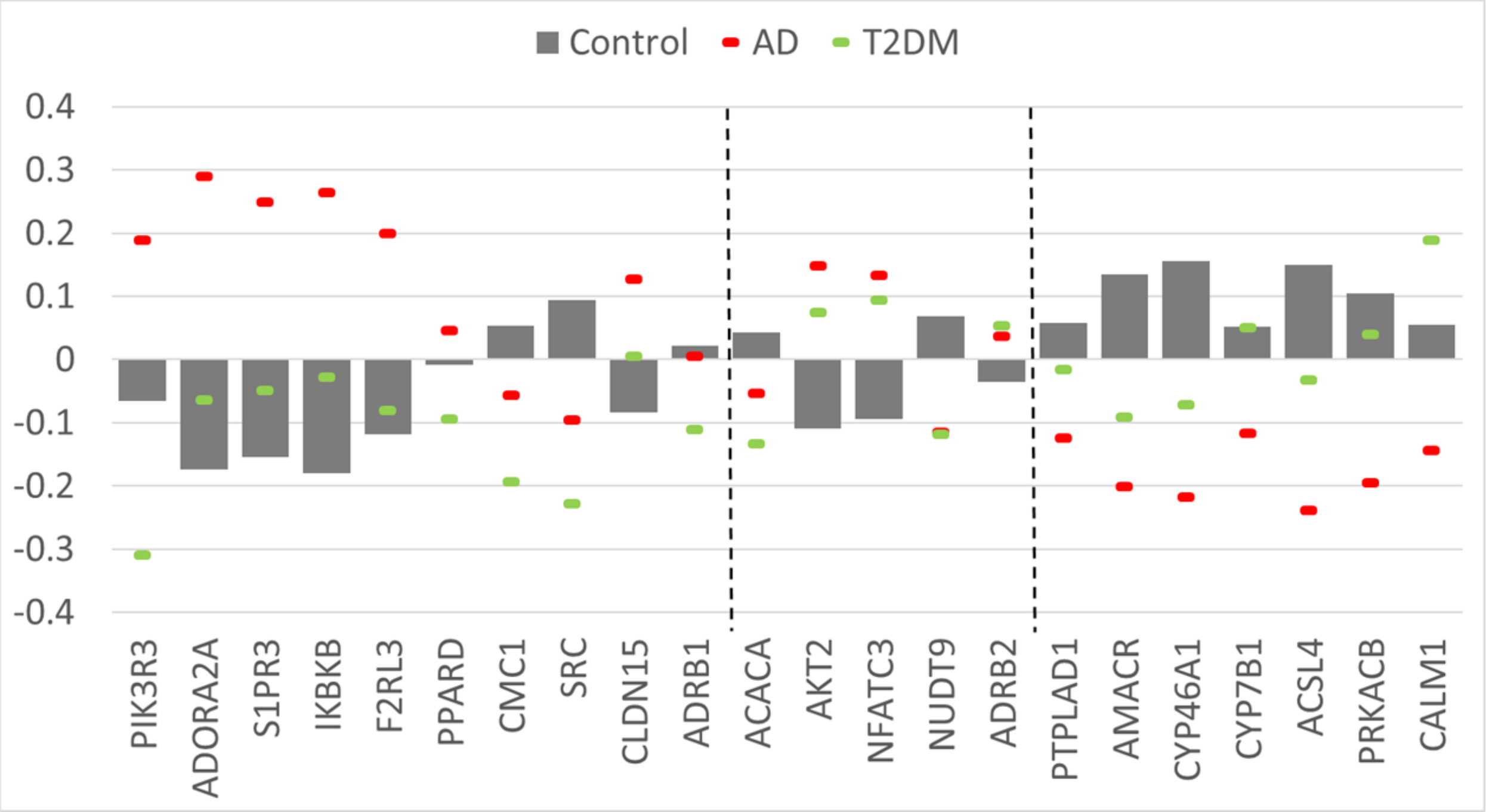

**Figure S3.** (A) Two major causal paths from APP to AD and T2DM in Figure 4; (B) Differential expressions of the genes along two major causal paths in (A) between AD and normal individuals where the nodes in red color represented over expressed genes, the nodes in green color represented the under expressed genes and the nodes in the grey color represented the genes showing no differential expressions; (C) Differential expressions of the genes along two major causal paths in (A) between T2DM and normal individuals where the nodes in red color represented over expressed genes, the nodes in green color represented the under expressed genes and the nodes in the grey color represented the genes showing no differential expressions.

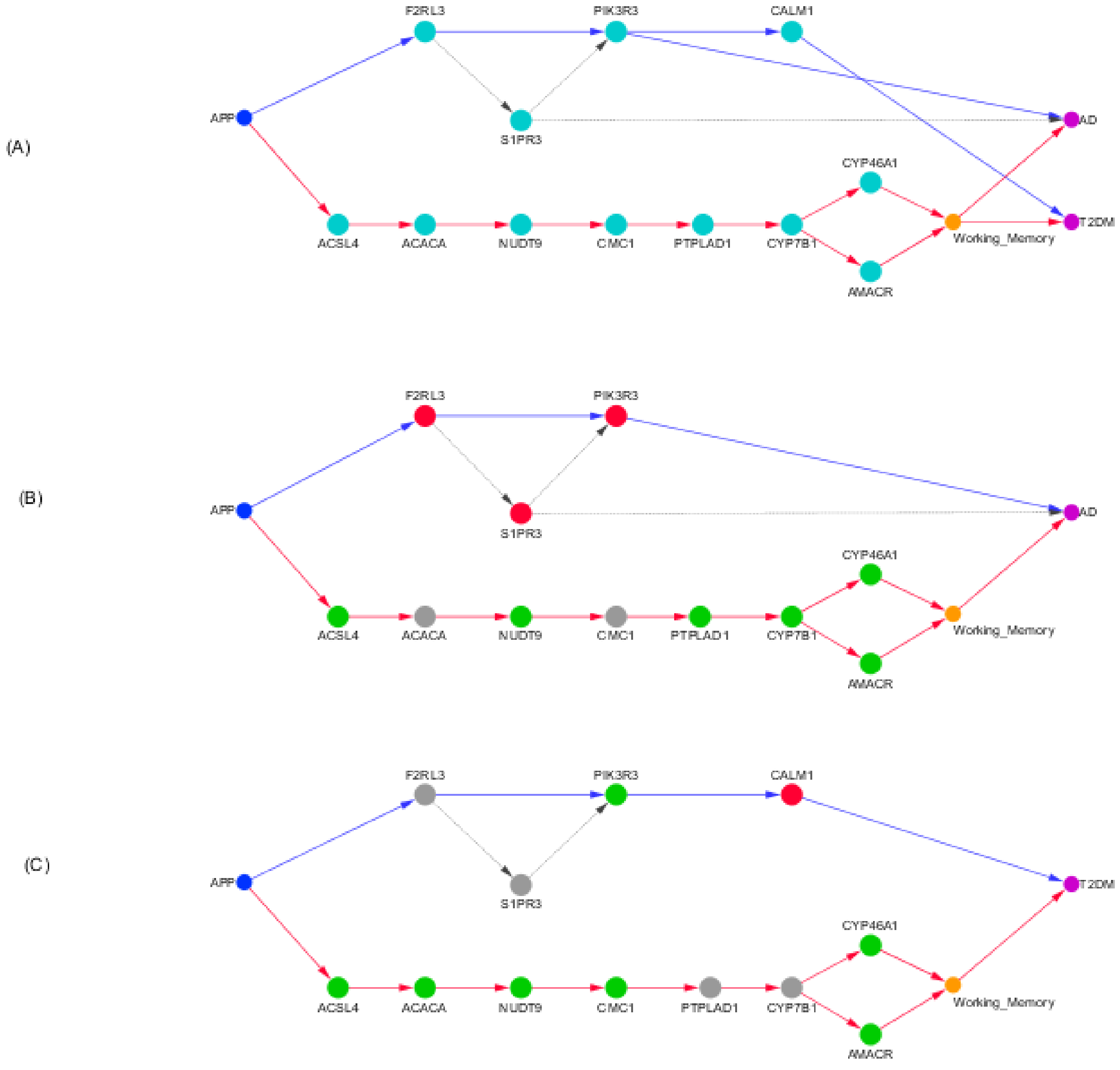

**Supplementary Table**

Table S1

| Gene | T2DM | | AD | |
|---|---|---|---|---|
| | Causation | Association | Causation | Association |
| GTF2H2C | <E-06 | 0.000156 | <E-06 | 0.075045 |
| NAIP | <E-06 | 0.000785 | <E-06 | 0.057792 |
| RN7SL9P | <E-06 | 0.000156 | <E-06 | 0.075045 |
| RP11-497H16.5 | <E-06 | 0.001671 | <E-06 | 0.182268 |
| RP11-497H16.6 | <E-06 | 0.000334 | <E-06 | 0.200074 |
| SERF1B | <E-06 | 0.000447 | <E-06 | 0.163819 |
| SMN2 | <E-06 | 0.000447 | <E-06 | 0.163819 |
| ZNF658B | <E-06 | 0.867831 | <E-06 | 0.102522 |

Table S2

| | AD | | T2DM | |
|---|---|---|---|---|
| | Causation | Association | Causation | Association |
| LINC01123 | <E-06 | 0.209055802 | <E-06 | 0.007821734 |
| ZBTB45P1 | <E-06 | 0.209055802 | <E-06 | 0.007821734 |
| GTF2H2C | <E-06 | 0.045337348 | <E-06 | 0.000155933 |
| NAIP | <E-06 | 0.023560082 | <E-06 | 0.000785176 |
| RN7SL9P | <E-06 | 0.075045141 | <E-06 | 0.000155933 |
| RP11-497H16.5 | <E-06 | 0.027245868 | <E-06 | 0.001670784 |
| RP11-497H16.6 | <E-06 | 0.106195835 | <E-06 | 0.000334017 |
| SERF1B | <E-06 | 0.017756029 | <E-06 | 0.000446698 |
| SMN2 | <E-06 | 0.017756029 | <E-06 | 0.000446698 |
| RN7SL763P | <E-06 | 0.242608501 | <E-06 | 0.887587122 |
| RANP9 | <E-06 | 0.153098533 | <E-06 | 0.206418787 |
| ZNF658B | <E-06 | 0.006744952 | <E-06 | 0.203671552 |
| AC132872.2 | <E-06 | 0.118907963 | <E-06 | 0.047484691 |

Table S3

| | AD | | T2DM | |
|---|---|---|---|---|
| | Causation | Association | Causation | Association |
| IGKV6D-21 | <E-06 | 0.00936 | | 0.00838 |
| RN7SL632P | <E-06 | 0.00053 | | 0.10824 |
| CDH18 | <E-06 | 0.00736 | | 0.11987 |
| HLA-DRB5 | <E-06 | 0.01958 | | 0.01317 |
| RPL3P2 | <E-06 | 0.00391 | | 0.05737 |
| FAM74A6 | <E-06 | 0.00256 | | 0.17522 |
| RNU6-156P | <E-06 | 0.00209 | | 0.00086 |
| RP11-15J10.1 | <E-06 | 0.00876 | | 0.01246 |
| RP11-262H14.4 | <E-06 | 0.00173 | | 0.01412 |
| RP11-318K12.2 | <E-06 | 0.0031 | | 0.00086 |
| RNU6-702P | <E-06 | 0.01116 | | 0.26546 |
| AL021920.1 | <E-06 | 0.00529 | | 0.10803 |
| AL590452.1 | <E-06 | 0.0034 | | 0.01512 |
| EIF1AXP1 | <E-06 | 0.00529 | | 0.10803 |
| FAM231B | <E-06 | 0.00187 | | 0.28261 |
| PRAMEF11 | <E-06 | 0.00122 | | 0.07388 |
| RP5-845O24.8 | <E-06 | 0.00122 | | 0.07388 |
| TTC3 | <E-06 | 0.0102 | | 0.01343 |
| AC009237.16 | <E-06 | 0.0024 | | 0.17164 |
| AC009237.17 | <E-06 | 0.0024 | | 0.17164 |

Table S4

| Gene | P-value | | | |
|---|---|---|---|---|
| | Causation | | Association | |
| | AD | T2DM | AD | T2DM |
| PLA2G15 | <0.0001 | 0.6633 | 0.0012 | 0.7645 |
| GRPEL2 | <0.0001 | 0.3611 | 0.0026 | 0.9894 |
| TCEAL6 | <0.0001 | 0.7324 | 0.0001 | 0.3729 |
| C19orf71 | <0.0001 | 0.8063 | 0.0002 | 0.4285 |
| GRTP1 | <0.0001 | 0.7326 | 0.0004 | 0.4044 |
| CTC-471C19.1 | <0.0001 | 0.9039 | <0.0001 | 0.8677 |
| GAP43 | <0.0001 | 0.413 | 0.0001 | 0.2183 |
| NTNG1 | <0.0001 | 0.2216 | <0.0001 | 0.298 |
| BIN1 | <0.0001 | 0.9583 | 0.4087 | 0.809 |
| ZNF683 | <0.0001 | 0.5699 | 0.0098 | 0.391 |
| PIGS | <0.0001 | 0.361 | 0.0624 | 0.7423 |
| MRPS18A | <0.0001 | 0.4478 | 0.0004 | 0.8081 |
| PRKCD | 0.0001 | 0.5847 | 0.0006 | 0.0814 |
| FNDC5 | 0.0001 | 0.8283 | 0.0000 | 0.9253 |
| SLC7A14 | 0.0001 | 0.9696 | 0.0101 | 0.7183 |
| GRAMD1B | 0.0001 | 0.0005 | 0.0171 | 0.6413 |
| USP12 | 0.0001 | 0.5937 | 0.041 | 0.0269 |
| DHRS7 | 0.0001 | 0.9111 | 0.0571 | 0.4734 |
| OAZ1 | 0.0001 | 0.7556 | 0.0089 | 0.6632 |

Table S5

| Gene | P-value | | | |
|---|---|---|---|---|
| | Causation | | Association | |
| | T2DM | AD | T2DM | AD |
| GRN | <0.0001 | 0.3019 | <0.0001 | 0.3037 |
| ATP1B3 | <0.0001 | 0.3848 | <0.0001 | 0.3358 |
| CTD-3065J16.6 | <0.0001 | 0.5351 | 0.4403 | 0.2133 |
| BX571672.1 | <0.0001 | 0.587 | 0.5182 | 0.3088 |
| SLC25A35 | <0.0001 | 0.7881 | 0.6975 | 0.3619 |
| PRR12 | <0.0001 | 0.9786 | 0.3704 | 0.281 |
| AOC3 | <0.0001 | 0.9936 | 0.1873 | 0.0678 |

Table S6

| Methylated Gene | P-value | |
|---|---|---|
| | Causation | Association |
| EMX2OS | < 1e-4 | 0.1973 |
| PIPOX | < 1e-4 | 0.6245 |
| DHX8 | < 1e-4 | 0.7317 |
| cg04413644 | < 1e-4 | 0.0086 |
| cg12049093 | < 1e-4 | 0.1122 |
| cg21346589 | < 1e-4 | 0.235 |
| cg18527583 | < 1e-4 | 0.0126 |
| cg00639635 | < 1e-4 | 0.2032 |
| cg22133973 | < 1e-4 | 0.0004 |
| cg12644659 | < 1e-4 | 0.7813 |
| cg12949927 | < 1e-4 | 0.0915 |
| cg20714487 | < 1e-4 | 0.0000 |
| cg03701930 | < 1e-4 | 0.5572 |
| CARKD | < 1e-4 | 0.2309 |
| NYNRIN | < 1e-4 | 0.1213 |
| IL2RA | < 1e-4 | 0.8899 |
| cg02438164 | < 1e-4 | 0.0029 |

Table S7

| Methylated Gene | P-value | |
|---|---|---|
| | Causation | Association |
| cg27424148 | < 1e-4 | 0.8752 |
| MIR220B | < 1e-4 | 0.3568 |
| cg16991316 | < 1e-4 | 0.0292 |
| cg01433468 | < 1e-4 | 0.4165 |
| cg04124260 | < 1e-4 | 0.0587 |
| cg06769739 | < 1e-4 | 0.0004 |
| cg11989330 | < 1e-4 | 0.0006 |
| cg17883371 | < 1e-4 | 0.0075 |
| GSTTP1 | < 1e-4 | 0.0016 |
| YWHAQ | < 1e-4 | 0.6393 |
| cg25004193 | < 1e-4 | 0.0001 |
| cg25757820 | < 1e-4 | 0.2169 |
| cg02093808 | < 1e-4 | 0.2378 |
| C11orf45 | < 1e-4 | 0.1566 |
| cg00850073 | < 1e-4 | 0.0136 |
| NME5 | < 1e-4 | 0.1576 |
| cg14727987 | < 1e-4 | 0.0011 |
| cg10245123 | < 1e-4 | 0.0014 |
| MFHAS1 | < 1e-4 | 0.5626 |
| cg11388673 | < 1e-4 | 0.6857 |
| cg18001780 | < 1e-4 | 0.0127 |
| KRT77 | < 1e-4 | 0.0291 |
| cg16624888 | < 1e-4 | 0.0000 |
| TPT1 | < 1e-4 | 0.2871 |
| CCDC70 | < 1e-4 | 0.0186 |
| cg01358406 | < 1e-4 | 0.0119 |
| cg03608502 | < 1e-4 | 0.1001 |

Table S8. A list of 120 DNA methylation sites/genes that were indirectly connected to AD and T2DM.

| Methylation Site/Gene | Expressed Genes | P-value | | Reference Methylation Site in the Gene |
|---|---|---|---|---|
| | | Causation | Association | |
| cg18747197 | QKI | 0.00004 | 4.04E-06 | |
| cg19351026 | GABRA2 | 0.00018 | 1.01E-05 | |
| cg27665808 | snoU13 | 0.00014 | 1.17E-05 | |
| GSTTP1 | RP11-259G18.2 | 0.00007 | 1.55E-05 | cg10678937,cg11141652,cg15242686,cg22666875 |
| cg12363375 | CTD-3025N20.2 | 0.00021 | 1.98E-05 | |
| cg16677162 | GPX6 | 0.00011 | 2.24E-05 | |
| cg13401079 | RCOR1 | 0.00015 | 2.77E-05 | |
| cg17670237 | RP11-259G18.2 | 0.00015 | 2.77E-05 | |
| cg07536144 | U1 | 0.00001 | 4.58E-05 | |
| ABCA10 | RP11-86H7.7 | 0.0001 | 4.64E-05 | cg13849142,cg14019757,cg14069205 |
| cg22463795 | GABRA2 | 0.00019 | 4.65E-05 | |
| cg01797450 | ASS1 | 8.00E-05 | 5.03E-05 | |
| cg23454003 | PSMB6 | 9.00E-05 | 5.45E-05 | |
| cg05455747 | ZC3H4 | 0.00E+00 | 6.23E-05 | |
| cg17526301 | GPX6 | 3.00E-05 | 8.06E-05 | |
| cg10061805 | PPIAP2 | 1.80E-04 | 1.53E-04 | |
| cg02190400 | ZC3H4 | 5.00E-05 | 1.66E-04 | |
| cg15304404 | RXFP4 | 2.80E-04 | 1.83E-04 | |
| cg24751928 | CTD-2501E16.2 | 4.00E-05 | 1.96E-04 | |
| cg17125990 | GABRA2 | 3.00E-05 | 2.06E-04 | |

| Methylation Site/Gene | Expressed Genes | P-value | | Reference Methylation Site in the Gene |
|---|---|---|---|---|
| | | Causation | Association | |
| cg00415011 | GARNL3 | 0.00E+00 | 2.08E-04 | |
| cg23340218 | WNT3A | 3.00E-05 | 2.45E-04 | |
| cg26274929 | UTP6 | 5.00E-05 | 2.66E-04 | |
| cg09977969 | CTD-2116F7.1 | 2.00E-04 | 2.68E-04 | |
| cg02461269 | CTD-3025N20.2 | 6.00E-05 | 3.15E-04 | |
| cg02523270 | GARNL3 | 3.60E-04 | 3.24E-04 | |
| cg09550810 | GARNL3 | 2.00E-05 | 3.80E-04 | |
| cg07475973 | RP11-265D19.6 | 1.20E-04 | 3.91E-04 | |
| cg21686890 | ABCB6 | 2.00E-05 | 3.98E-04 | |
| cg21099148 | snoU13 | 1.90E-04 | 4.26E-04 | |
| cg01870681 | GPX6 | 1.90E-04 | 4.40E-04 | |
| cg25112877 | WNT3A | 1.40E-04 | 4.63E-04 | |
| cg19711815 | RP11-259G18.2 | 8.00E-05 | 5.49E-04 | |
| cg06447341 | TTC37 | 3.00E-05 | 6.86E-04 | |
| ch.2.3048096R | ASS1 | 2.00E-05 | 7.17E-04 | |
| cg07536144 | PLCH1-AS1 | 9.00E-05 | 7.44E-04 | |
| cg15426660 | snoU13 | 1.00E-05 | 7.61E-04 | |
| cg08209099 | RP11-259G18.2 | 7.00E-05 | 7.85E-04 | |
| cg22891595 | GARNL3 | 1.00E-05 | 8.71E-04 | |
| cg06181286 | CBLN4 | 0.00E+00 | 9.85E-04 | |
| cg14739859 | RP11-259G18.2 | 5.00E-05 | 9.89E-04 | |
| cg12548341 | GNAL | 1.30E-04 | 1.06E-03 | |

| Methylation Site/Gene | Expressed Genes | P-value | | Reference Methylation Site in the Gene |
|---|---|---|---|---|
| | | Causation | Association | |
| cg15426660 | SNTB2 | 9.00E-05 | 1.16E-03 | |
| cg10059756 | GABRA2 | 4.00E-05 | 1.21E-03 | |
| cg21205865 | snoU13 | 4.00E-05 | 1.32E-03 | |
| cg00039801 | RP11-259G18.2 | 6.00E-05 | 1.35E-03 | |
| cg10316834 | snoU13 | 1.00E-04 | 1.35E-03 | |
| cg20078879 | RPS4X | 9.00E-05 | 1.37E-03 | |
| cg24015081 | CTD-2176I21.1 | 1.80E-04 | 1.45E-03 | |
| cg25214310 | SUGT1 | 3.70E-04 | 1.53E-03 | |
| cg14687029 | ABCB6 | 8.00E-05 | 1.56E-03 | |
| cg05157340 | RANBP2 | 1.40E-04 | 2.00E-03 | |
| cg06709828 | CTD-3025N20.2 | 0.00E+00 | 2.08E-03 | |
| cg06112654 | SNTB2 | 4.00E-05 | 2.09E-03 | |
| cg06112654 | snoU13 | 6.00E-05 | 2.18E-03 | |
| cg22651787 | FAM188B | 5.00E-05 | 2.50E-03 | |
| cg14577406 | RBPMS2 | 0.00E+00 | 2.58E-03 | |
| cg04854637 | SUGT1 | 3.00E-05 | 3.27E-03 | |
| cg12180270 | GABRA2 | 2.00E-05 | 3.31E-03 | |
| cg04606076 | RXFP4 | 1.20E-04 | 3.32E-03 | |
| cg02486332 | CASP12 | 1.20E-04 | 3.43E-03 | |
| cg11868247 | RP11-351K16.4 | 7.00E-05 | 3.50E-03 | |
| cg13001963 | RXFP4 | 0.00E+00 | 3.99E-03 | |
| cg24430419 | snoU13 | 1.00E-04 | 4.03E-03 | |

| Methylation Site/Gene | Expressed Genes | P-value | | Reference Methylation Site in the Gene |
|---|---|---|---|---|
| | | Causation | Association | |
| cg07716131 | SAA1 | 1.60E-04 | 4.57E-03 | |
| MIR145 | RCOR1 | 3.00E-05 | 4.80E-03 | cg01310120,cg08537847,cg11671363,cg22941668,cg23917868,cg27083040 |
| cg16101962 | GARNL3 | 2.00E-05 | 4.95E-03 | |
| cg10238080 | snoU13 | 6.30E-04 | 5.01E-03 | |
| cg12094552 | SNTB2 | 5.00E-05 | 5.25E-03 | |
| cg23346544 | ARHGAP20 | 0.00E+00 | 5.56E-03 | |
| cg13332807 | PKN2 | 1.30E-04 | 6.11E-03 | |
| TDGF1 | ST6GALNAC6 | 5.00E-05 | 7.25E-03 | cg06174858 |
| cg24617444 | RUSC1 | 0.00E+00 | 7.76E-03 | |
| ANKRD53 | CPE | 2.10E-04 | 7.92E-03 | cg00421335,cg01154254,cg04076766,cg04737087,cg05472974,cg05492660,cg06783668,cg07903989,cg12428298,cg15165122,cg18006568,cg18313051,cg19111030,cg19244342,cg19490001,cg20814026,cg22050950,cg23060872,cg24573743,cg27665449 |
| cg20918393 | PPIAP2 | 9.00E-05 | 8.20E-03 | |
| cg23925513 | RARB | 1.20E-04 | 8.65E-03 | |
| cg18105749 | BRSK2 | 5.00E-05 | 9.15E-03 | |
| cg16624888 | RXFP4 | 6.00E-05 | 9.48E-03 | |
| cg09315586 | SUGT1 | 1.00E-05 | 9.81E-03 | |
| cg20078879 | DNAJC1 | 7.00E-05 | 1.01E-02 | |

| Methylation Site/Gene | Expressed Genes | P-value | | Reference Methylation Site in the Gene |
|---|---|---|---|---|
| | | Causation | Association | |
| cg00651087 | ZNF114 | 5.00E-05 | 1.12E-02 | |
| cg04606076 | C2CD2L | 1.50E-04 | 1.22E-02 | |
| cg10061805 | FAN1 | 1.90E-04 | 1.25E-02 | |
| cg25900943 | SNTB2 | 1.00E-05 | 1.31E-02 | |
| HRCT1 | DCTN2 | 2.80E-04 | 1.31E-02 | cg02258201,cg05470166 |
| cg05521150 | PMS2CL | 5.00E-05 | 1.44E-02 | |
| cg11868247 | CSNK1G1 | 7.00E-05 | 1.46E-02 | |
| cg11868247 | EIF6 | 6.00E-05 | 1.66E-02 | |
| cg24631482 | LA16c-444G7.2 | 0.00005 | 1.77E-02 | |
| cg04858776 | OR2L13 | 0.00015 | 1.86E-02 | |
| cg15022039 | SRD5A1 | 0.00033 | 1.90E-02 | |
| C6orf223 | SMO | 0.00003 | 2.01E-02 | cg02213045,cg04386144,cg09745336,cg10938046,cg11201894,cg13900100,cg15629096,cg16039071,cg18949415,cg19274606,cg24630373,cg25360181 |
| cg17624891 | DGKK | 0.00007 | 2.04E-02 | |
| cg25816127 | WNT3A | 0 | 2.08E-02 | |
| cg26686150 | FAM213B | 0.00001 | 2.14E-02 | |
| cg19937979 | hsa-mir-146a | 0.00002 | 2.15E-02 | |
| cg04819760 | FAH | 0.00057 | 2.28E-02 | |
| cg07929768 | POTEG | 1.60E-04 | 2.33E-02 | |
| cg26095395 | ADRA1B | 0.00E+00 | 2.54E-02 | |

| Methylation Site/Gene | Expressed Genes | P-value | | Reference Methylation Site in the Gene |
|---|---|---|---|---|
| | | Causation | Association | |
| cg14145524 | XRCC5 | 0.00E+00 | 2.56E-02 | |
| cg17214023 | INPP5F | 8.00E-05 | 2.59E-02 | |
| cg16609139 | SDS | 1.30E-04 | 2.74E-02 | |
| ch.11.96774805R | RPS4X | 3.00E-05 | 2.79E-02 | |
| ch.10.109266902R | RP11-351K16.4 | 0.00013 | 2.95E-02 | |
| cg02796279 | OR7C1 | 0.00004 | 3.29E-02 | |
| cg19937979 | NDUFA8 | 0.00003 | 3.42E-02 | |
| ch.2.1894803R | NUMB | 0.00006 | 3.44E-02 | |
| cg14257429 | SNTB2 | 0.00002 | 3.66E-02 | |
| cg08122831 | RP11-284N8.3 | 0.00001 | 3.78E-02 | |
| cg11465213 | EXOSC2 | 0.00006 | 4.05E-02 | |
| cg05766107 | ADCYAP1R1 | 0.00005 | 4.08E-02 | |
| cg00895132 | ZNF233 | 0.00003 | 4.19E-02 | |
| cg19355069 | ESD | 0.00014 | 4.32E-02 | |
| cg18978531 | C2orf18 | 0.00009 | 4.72E-02 | |
| MYCT1 | RP11-265D19.6 | 0.00002 | 4.76E-02 | cg02830467,cg15961007 |
| cg03257930 | FADS6 | 0.00003 | 4.81E-02 | |
| cg11540476 | RP11-114F3.5 | 0.00001 | 4.86E-02 | |